\documentclass[aps,prb,twocolumn,superscriptaddress]{revtex4-2}
\usepackage{physics}
\usepackage{graphicx}
\usepackage{lineno,hyperref}
\usepackage{amsmath,amssymb}
\usepackage{array}

\hypersetup{
    colorlinks=true,
    linkcolor=blue,
    filecolor=magenta,      
    citecolor=blue,
    urlcolor=magenta,
}

\newcommand{\bk}{\mathbf{k}}
\newcommand{\br}{\mathbf{r}}
\newcommand{\bQ}{\mathbf{Q}}

\begin{document}
\title{Strong-coupling expansion of multi-band interacting models: mapping onto the transverse-field $J_1$-$J_2$ Ising model}

\date{\today}
\author{Xiaoyu Wang}
\affiliation{National High Magnetic Field Laboratory, Tallahassee, FL 32310, USA}
\author{Morten H. Christensen}
\affiliation{Niels Bohr Institute, University of Copenhagen, Denmark}
\author{Erez Berg}
\affiliation{Department of Condensed Matter Physics, Weizmann Institute of Science, Rehovot 76100, Israel}
\author{Rafael M. Fernandes}
\affiliation{School of Physics and Astronomy, University of Minnesota, Minneapolis, MN 55455, USA}

\begin{abstract}
   We investigate a class of two-dimensional two-band microscopic models in which the inter-band repulsive interactions play the dominant role. We first demonstrate three different schemes of constraining the ratios between the three types of inter-band interactions -- density-density, spin exchange, and pair-hopping -- that render the model free of the fermionic sign-problem for any filling and, consequently, amenable to efficient Quantum Monte Carlo simulations. We then study the behavior of these sign-problem-free models in the strong-coupling regime. In the cases where spin-rotational invariance is preserved or lowered to a planar symmetry, the strong-coupling ground state is a quantum paramagnet. However, in the case where there is only a residual Ising symmetry, the strong-coupling expansion maps onto the transverse-field $J_1$-$J_2$ Ising model, whose pseudospins are associated with local inter-band magnetic order. We show that by varying the band structure parameters within a reasonable range of values, a variety of ground states and quantum critical points can be accessed in the strong-coupling regime, some of which are not realized in the weak-coupling regime. We compare these results with the case of the single-band Hubbard model, where only intra-band repulsion is present, and whose strong-coupling behavior is captured by a simple Heisenberg model.
\end{abstract}

\maketitle

\section{Introduction}

In systems of interacting electrons, a strong Coulomb repulsion can give rise to correlated insulating states in which charge carriers become localized. The low-energy properties are then usually determined by emergent charge-neutral excitations. A prime example is the Mott-insulating phase observed in several transition metal oxides, including the parent compounds of the high-temperature cuprate superconductors~\cite{Anderson1987,Imada_review,Lee2006}. The low energy excitations are typically magnetic in nature, since the local spins usually order at low enough temperatures inside the Mott state. More exotic correlated phenomena may arise in systems where the electrons have additional degrees of freedom besides spin. Examples include the orbital-selective Mott transition~\cite{Medici2014,Yu2013,Dagotto2014} and the Hund's metallic state~\cite{Haule2009,Georges2013} in multi-orbital systems, as well as ferromagnetic and Chern insulating phases in graphene-based systems with valley degrees of freedom \cite{Kang2019,TBGIV2020,Bultinck2020,Meng2021}.

From a theoretical perspective, while these materials have important structural and chemical differences, it is useful to consider simple models that may capture universal emergent behaviors associated with these correlated phases~\cite{Anderson1972}. In this regard, the Hubbard model~\cite{Hubbard1963,Gutzwiller1963,Kanamori1963} is certainly among the most studied models in condensed matter physics, consisting of a kinetic hopping term (with coefficient $t$) and an onsite Hubbard repulsion term (with coefficient $U$) involving electrons on a single orbital. Upon increasing $U$, one generally expects a metal to Mott-insulator transition~\cite{Georges1996}. However, this is not the full story. As Phil Anderson showed in Ref.~\onlinecite{Anderson_superexchange}, a perturbative calculation in the strong-coupling regime ($U \gg t$) reveals that the spins of the localized charge carriers experience a superexchange interaction promoted by virtual hopping processes. As a result, at half-filling, the insulating state is expected to display long-range magnetic N\'eel order -- unless frustration is present, in which case spin liquid phases may appear~\cite{Anderson1987}. Moving away from half-filling, one obtains the rich $t$-$J$ model \cite{Lee2006}. Interestingly, in the weak-coupling regime ($U \ll t$), perturbative calculations generically find a N\'eel state at half-filling ~\cite{Schrieffer1989,Schulz1987}, which can be a metal or a Slater insulator depending on additional hopping parameters. As a result, the magnetic order in the weak-coupling and strong-coupling regimes are the same. 

For intermediate coupling strengths and away from half-filling, unconventional superconductivity is generally expected from both strong-coupling ~\cite{Anderson1987,Kotliar1988} and weak-coupling perspectives ~\cite{Scalapino1986,Monthoux1991,Chubukov_Pines_Schmalian}. Assessing this regime of moderate correlations, however, is theoretically challenging due to its non-perturbative nature. Numerical methods have played an important role in bridging the gap between the strong- and weak-coupling regimes of the Hubbard model ~\cite{LeBlanc2015}. These include density-matrix renormalization group methods~\cite{White1992,Noack1994, Jiang2019}, dynamical mean-field theory~\cite{Georges1996, Park2008,Gull2008,Dong2020}, and Quantum Monte Carlo (QMC) simulations \cite{Blankenbecler1981,White1989,Berg2012}. While the latter is a powerful method, due to its exact and unbiased nature, it suffers from the infamous fermionic sign problem, which, in the case of the Hubbard model, can only be avoided exactly at half filling and for bipartite lattices ~\cite{Hirsch1985}.

One of the main motivations to study the Hubbard model has undoubtedly been the cuprate high-temperature superconductors. While questions remain about whether this single-band model can capture their rich phenomenology ~\cite{Emery1987,Varma1997}, the rise of iron-based superconductors, ruthenates, and other multi-band systems has revived the interest in multi-orbital generalizations of the Hubbard model, the so-called Hubbard-Kanamori models (see, for example, Ref.~\onlinecite{Castellani1978}).
Several works have unearthed the unique properties of these models, including the importance of the Hund's rule coupling in promoting strong-coupling behavior~\cite{Haule2009,Georges1996} and the rich interplay between nesting-driven spin-density wave and unconventional superconductivity~\cite{Chubukov2008}. Strong-coupling expansions involving both spin and orbital degrees of freedom have also been widely employed \cite{Kruger2009}, resulting in complex Kugel-Khomskii effective models~\cite{Kugel_Khomskii}. 
From a numerical perspective, the fact that the Hubbard model suffers from the fermionic sign-problem may discourage the use of QMC methods to investigate the more complicated Hubbard-Kanamori models. However, as realized in Ref.~\onlinecite{Berg2012}, by extending the number of bands of certain low-energy models that describe electrons interacting with bosonic excitations, it is possible to completely avoid the sign problem due to the emergence of an anti-unitary symmetry~\cite{SCZhang2005}. A similar reasoning was put forward previously in Ref.~\onlinecite{Motome1997} in the context of the Hubbard-Kanamori model. Therefore, multi-band interacting models may provide a unique window into the regime of moderate correlations that is usually difficult to access in single-band models.

\begin{figure}
    \centering
    \includegraphics[width=\linewidth]{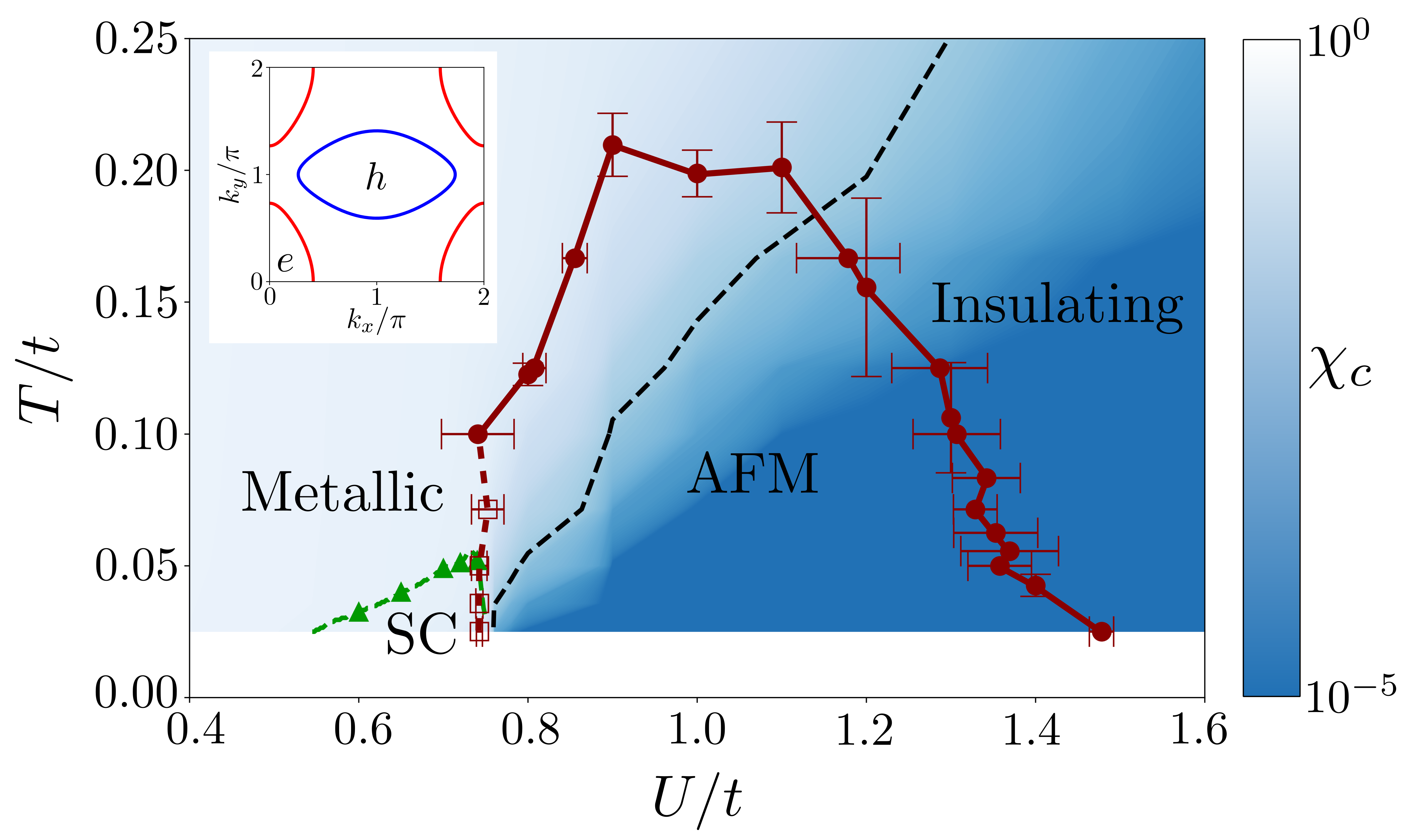}
    \caption{Phase diagram of the two-band electronic model with inter-band only repulsion, see Eq.~\eqref{eq:AttractiveSSIsing}. The dashed black line denotes a metal-to-insulator transition/crossover, whereas the green lines denote the superconducting (SC) dome. The dashed red line marks a first-order antiferromagnetic (AFM) phase transition. The AFM phase has a dome-like structure bounded by two putative quantum phase transitions. The inset depicts the Fermi surface of the non-interacting Hamiltonian. The color code refers to the charge compressibility $\chi_c$. Figure reproduced from Ref.~\onlinecite{Christensen2020}. Copyright 2020 by the American Physical Society.}
    \label{fig:IsingPhaseDiagram}
\end{figure}

In a previous work~\cite{Christensen2020}, together with Y. Schattner, we showed that a particular realization of the Hubbard-Kanamori model, formulated in band space rather than in orbital space and with spin-anisotropic interactions, is amenable to be simulated with a sign-problem-free QMC method. The key point is to set the intra-band repulsion to zero and consider only inter-band repulsion terms. While the opposite limit of large intra-band repulsion and vanishing inter-band terms should map directly onto the usual Hubbard model, the inter-band dominated regime that we considered in Ref.~\onlinecite{Christensen2020} has been little explored. As such, it has the potential to provide important insights onto the properties of multi-band interacting systems. Importantly, such a limit is not as artificial as it may first look: a weak-coupling renormalization-group (RG) analysis of this model shows that the inter-band terms grow much faster than the intra-band ones under the RG flow, in the case of a nearly-nested Fermi surface~\cite{Chubukov2008}. Moreover, the model remains sign-problem free even away from half-filling and for longer-range hopping parameters.

In Fig.~\ref{fig:IsingPhaseDiagram}, we reproduce the temperature-interaction strength phase diagram obtained by us in Ref.~\onlinecite{Christensen2020},
via QMC simulations for a model with nearly nested bands (see the inset).
Despite the absence of intra-band repulsion, a metal-to-insulator crossover takes place for intermediate coupling strengths (dashed black line). Superconductivity (SC, green line) is also found in the metallic state near the onset of N\'eel antiferromagnetic (AFM) order. The main feature of this phase diagram, however, is the emergence of an AFM dome (red line). Starting from the weak-coupling limit, it is not surprising that AFM order is only seen after the interaction strength overcomes a threshold value, since the nesting between the two bands is not perfect. What is more surprising is the apparent lack of AFM order in the strong-coupling limit, since N\'eel order is a hallmark of the Mott insulating state of the standard Hubbard model. In Ref.~\cite{Christensen2020}, we performed a strong-coupling expansion of this inter-band interacting model, and found the ground state to be in the quantum paramagnetic phase of an effective transverse-field Ising model, in agreement with the QMC results.

In this paper, we perform a strong-coupling expansion of the inter-band interacting model for an arbitrary band dispersion and for different types of spin-anisotropic interactions, thereby extending our previous results. In the case of inter-band interactions that preserve the SU$(2)$ spin rotational symmetry or lower it to a planar symmetry, we find a non-degenerate ground state for the single site problem, indicating a ``trivial" quantum paramagnetic ground state in the strong-coupling regime. On the other hand, for the case of inter-band interactions that lower the SU$(2)$ symmetry to an Ising symmetry, the ground state of the single-site problem is doubly degenerate. This gives rise to a pseudospin that corresponds to the two polarizations of the inter-band Ising-like magnetization. 
In terms of this pseudospin, the strong-coupling expansion of our microscopic interacting model maps onto a transverse-field Ising model with extended exchange interactions, i.e. nearest-neighbor interactions, next-nearest-neighbor interactions, etc. In particular, we show that, by changing the band dispersion parameters across a reasonable range of values, one can in principle access the entire phase diagram of the transverse-field $J_1$-$J_2$ Ising model. This includes the regime where the critical behavior is that of the simple transverse-field Ising model -- as was the case for the particular band parameters we considered in Ref.~\onlinecite{Christensen2020} -- or that of the transverse field Ashkin-Teller (four-state clock) model. We discuss how the latter can shed new light on the emergence of vestigial nematicity in systems that display stripe-type magnetic order, with possible implications for the coupled magnetic-nematic transitions of iron-based superconductors \cite{Fernandes2012}. Overall, our work reveals the richness of the strong-coupling regime of microscopic models in which inter-band interactions dominate, which can be quite different from the strong-coupling behavior of the standard intra-band dominated Hubbard model.    

Our paper is organized as follows: In Section \ref{sec:model} we introduce the microscopic two-band model, and show how restrictions placed on the interaction parameters lead to the emergence of an anti-unitary symmetry, which allows the model to be simulated using sign-problem free QMC. In Section \ref{sec:strong_exp} we perform a strong coupling expansion of the model with an inter-band spin-spin interaction of the Ising-type, and show that it is mapped onto the transverse-field $J_1$-$J_2$ Ising model. In Section \ref{sec:tfim}, we discuss the rich phase diagram of the transverse-field $J_1$-$J_2$ Ising model, as well as the choice of microscopic tight-binding parameters to achieve each ground state. Our conclusions are presented in Sec.~\ref{sec:conclusions}.

\section{Multi-band interacting model}\label{sec:model}
We start from a two-band electronic Hamiltonian with onsite interactions only:
\begin{equation}\label{eq:Chubukov2Band}
\begin{split}
    H & = H_2 + H_4,\\
    H_2  & =  \sum_{ij,\mu}\left[(t^c_{ij}-\mu^c\delta_{ij})c^\dagger_{i\mu}c_{j\mu}+(t^d_{ij}-\mu^d\delta_{ij})d^\dagger_{i\mu}d_{j\mu} \right],\\
     H_4 & = \sum_{i,\mu\nu} \left[ U_1 c^{\dagger}_{i\mu}c_{i\mu}d^{\dagger}_{i\nu}d_{i\nu} - U_2 c^{\dagger}_{i\mu}c_{i\nu}d^{\dagger}_{i\nu}d_{i\mu} \right.\\
     & + \frac{U_3}{2}\left(c^{\dagger}_{i\mu}c^\dagger_{i\nu}d_{i\nu}d_{i\mu}+h.c.\right) \\
    &\left. + U_4 c^{\dagger}_{i\mu}c_{i\mu}c^{\dagger}_{i\nu}c_{i\nu} + U_5 d^{\dagger}_{i\mu}d_{i\mu}d^{\dagger}_{i\nu}d_{i\nu}\right].
\end{split}
\end{equation}
Here, $c^{\dagger}_{i,\mu}$ ($d^{\dagger}_{i,\mu}$) creates an electron in band $c$ ($d$) at site $i$ of a square lattice with spin $\mu$. The non-interacting electronic dispersions $\varepsilon^c(\mathbf{k})$ and $\varepsilon^d(\mathbf{k})$ are obtained from the corresponding hopping parameters $t_{ij}^{c,d}$. The terms $\mu^c$ and $\mu^d$ contain implicitly both the chemical potential, given by $(\mu^c + \mu^d)/2$, and the onsite energies $\pm(\mu^c - \mu^d)/2$ associated with each band. There are five types of onsite electronic interactions, corresponding to intra-band density-density repulsion ($U_4$ and $U_5$), inter-band density-density repulsion $(U_1)$, spin exchange $(U_2)$ and pair hopping $(U_3)$. The model can been viewed as a projection of the usual two-orbital Hubbard-Kanamori model on the band basis~\cite{Chubukov2008}, with the assumption that the angle dependence of the projected interaction parameters can be neglected. 

Depending on the choice of band and interaction parameters, various low-energy phenomena can be explored. For example, if the non-interacting band dispersions are nearly nested (i.e. one hole-like Fermi pocket and another electron-like Fermi pocket of similar size and shape), density-wave order in different channels (i.e. spin, charge, loop-current) can be promoted at weak coupling depending on the values of the inter-band interactions $U_1$, $U_2$, and $U_3$ \cite{Chubukov2008}. Unconventional superconductivity, characterized by gaps of opposite signs on the two bands, is a close competitor of the density-wave order. On the other hand, if the intra-band interactions $U_4$ and $U_5$ are dominant, the model reduces to two nearly-independent copies of the single-band Hubbard model. Alternatively, by constraining the inter-band interactions  $U_1$, $U_2$, and $U_3$ to be twice the strength of the intra-band interactions, the model maps onto a two-layer Hubbard-model with nearest-neighbor hopping between the layers, which has been employed to study Cooper pairing due to an incipient band \cite{Maier2019}. 

As discussed in the introduction, our interest here is on the strong-coupling limit in the regime of dominant inter-band repulsion, as a counterpoint of the more widely studied regime of dominant intra-band repulsion. Therefore, hereafter we set $U_4=U_5=0$. An appealing feature of the model with inter-band interactions only is that, by properly setting the ratios between the three inter-band interactions, $U_1$, $U_2$, and $U_3$, the Hamiltonian can be efficiently solved numerically via sign-problem-free QMC simulations. This is because the interaction term can then be written as an effective attractive term in the inter-band spin-channel. Introducing the notation $\psi_i=(c_{i\uparrow},c_{i\downarrow},d_{i\uparrow},d_{i\downarrow})^T$, we define the inter-band spin order parameter as: $M^a_i \equiv \sum_{\alpha\beta}
 \psi^\dagger_{i\alpha}(\sigma^a\rho^x)_{\alpha\beta}\psi_{i\beta}$, where $\sigma$ and  $\rho$ are Pauli matrices acting in spin and band space respectively, and $(\alpha,\beta)$ are indices of the vector space spanned by $\psi_i$. 

There are three different possibilities to do such a rewriting of the interaction term. The first one corresponds to choosing $U_1=4U$, $U_2=2U$, and $U_3=6U$, which leads to:
\begin{equation}
    H_{4,\text{Heisenberg}} =  - U \sum_{a=x,y,z}\sum_{i}  M_i^a M_i^a \,. \label{eq:AttractiveSSHeisenberg}
\end{equation}
Here, the subscript ``Heisenberg" is used to emphasize the fact that the interaction term is invariant under spin SU(2) rotational symmetry, reminiscent of the Heisenberg model.  Other ratios between $U_1$, $U_2$, and $U_3$ allow for similar types of rewriting in terms of ``XY" and ``Ising" inter-band spins. Of course, in these cases the interactions must break spin-rotational symmetry. This is not an unreasonable assumption, since spin-orbit coupling naturally breaks spin-rotational symmetry in actual materials. In our treatment, such effects are treated on a phenomenological level only. In particular, setting  $U_1=4U(1-\delta_{\mu\nu})$, $U_2=0$, and $U_3=4U$ we obtain
\begin{equation}
    H_{4,\text{XY}} =  - U \sum_{a=x,y}\sum_{i}M_i^aM_i^a \,.\label{eq:AttractiveSSXY}
\end{equation}
Note that the interaction term possesses a residual $U(1)$ (or planar) symmetry, similar to the XY model. Finally, in the case where $U_1=4U\delta_{\mu\nu}$, $U_2=2U$, and $U_3=2U$, $H_4$ can be rewritten as
\begin{equation}
    H_{4,\text{Ising}} =  - U \sum_{i} M_i^zM_i^z\,, \label{eq:AttractiveSSIsing}
\end{equation}
where only a $\mathbb{Z}_2$ symmetry is preserved.

The key motivation to introduce these restrictions on the inter-band repulsions is that the resulting Hamiltonian can be simulated with QMC without the sign-problem. This can be seen by decoupling the quartic terms via a Hubbard-Stratonovich field $\{\phi_i^a(\tau)\}$, where $\tau$ denotes imaginary time. As a result, the partition function can be represented as 
\begin{equation}
    \mathcal{Z}=\int D[\phi] \det\{\widehat{G}^{-1}(\phi)\}\exp(-\mathcal{S}_{\phi})\,.
\end{equation}
Here $\mathcal{S}_{\phi}=\int_0^\beta \mathrm{d}\tau \frac{1}{4U}[\phi_i^a(\tau)]^2$ and $\widehat{G}\equiv (\partial_\tau + H_2+H^\phi_{2})^{-1}$ is the fermionic Green's function, with $H_{2}^{\phi}=\sum_i \phi_i^a(\tau)M^a_i$. In the single-particle Hilbert space, the quadratic Hamiltonian ($H_2+H_{2}^\phi$) has an anti-unitary symmetry $\widehat{U}=i\sigma^y\rho^zK$, where $K$ denotes complex conjugation. Since $\widehat{U}^2=-1$, the eigenstates of the single particle spectrum are always doubly degenerate (an analogue of a Kramer's doublet), which guarantees a positive fermionic determinant \cite{SCZhang2005}. In Ref. \cite{Christensen2020}, we used determinantal QMC to study the Ising case in Eq.~(\ref{eq:AttractiveSSIsing}), resulting in the phase diagram of Fig. \ref{fig:IsingPhaseDiagram} discussed in the Introduction.

It is interesting to note that the restriction on the ratios between $U_1$, $U_2$, and $U_3$ that is needed to ``eliminate" the fermionic sign-problem is a much milder constraint than the corresponding restriction on the single-band Hubbard model. In the latter case, one needs to impose half-filling and hopping parameters that preserve the bipartite nature of the square lattice. In the present case, on the other hand, there are no restrictions on the filling and on the type of hopping parameters. The reason is because the avoidance of the fermionic sign-problem arises from an anti-unitary symmetry related to the intrinsic two-band nature of the problem, in the same spirit as in Ref. \onlinecite{SCZhang2005}.

\section{Strong-coupling expansion}\label{sec:strong_exp}

While in our previous work \cite{Christensen2020} we performed a strong-coupling expansion of the Ising case [Eq.~(\ref{eq:AttractiveSSIsing})] for a specific tight-binding parameters set, here we explore the strong-coupling regime of the three cases presented in Eqs.~\eqref{eq:AttractiveSSHeisenberg}--\eqref{eq:AttractiveSSIsing} for an arbitrary tight-binding parametrization. The procedure consists of first neglecting the kinetic terms and then solving the interacting Hamiltonian exactly on a single site. Next, we treat the kinetic terms perturbatively, and discuss the strong-coupling physics using an effective Hamiltonian in terms of the available degrees of freedom ~\cite{SachdevQPT}. As we will show, the Ising case provides the only non-trivial ground state at strong-coupling, mapping onto the rich transverse-field $J_1$-$J_2$ Ising model. 

\subsection{Single-site exact solutions}

\begin{figure}
    \centering
    \includegraphics[width=\linewidth]{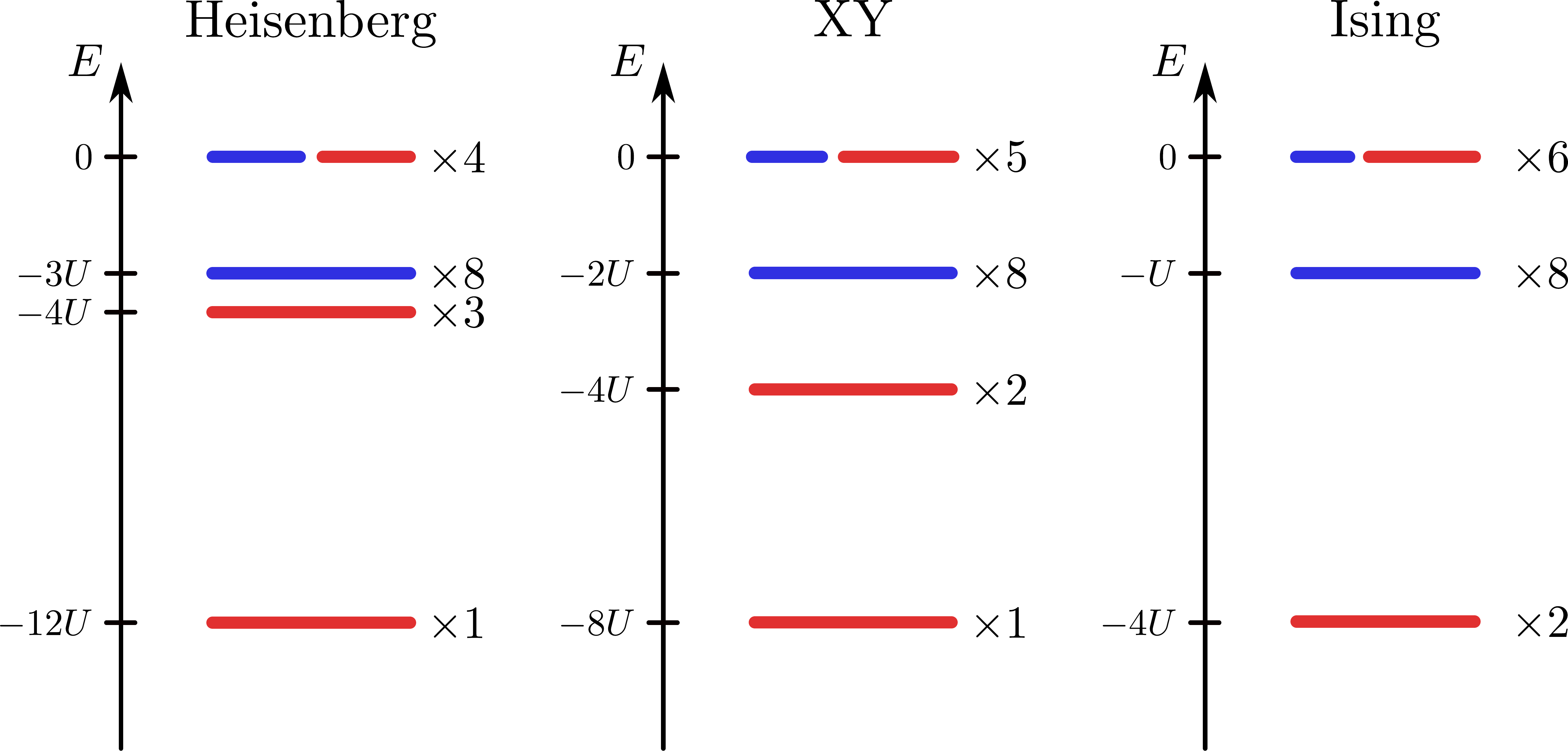}
    \caption{Energy spectrum and level degeneracy of the single-site interacting Hamiltonian $H_4$ for the Heisenberg [Eq.~\eqref{eq:AttractiveSSHeisenberg}], XY [Eq.~\eqref{eq:AttractiveSSXY}], and Ising [Eq.~\eqref{eq:AttractiveSSIsing}] cases. States in the $n=2$ filling sector are denoted in red.}
    \label{fig:single_site}
\end{figure}

The energy spectra of the interacting Hamiltonians in Eqs.~\eqref{eq:AttractiveSSHeisenberg}--\eqref{eq:AttractiveSSIsing} on a single site are illustrated in Figure~\ref{fig:single_site}. The local Fock space can be written as $\ket{\eta_c;\eta_d}$ where $\eta_c,\eta_d\in \{0,\uparrow,\downarrow,\uparrow\downarrow\}$. This is a 16-dimensional vector space. Due to number conservation, the Hamiltonians can be analyzed within any given integer electron filling $n=0, \dots, 4$. The linear dimension for each filling factor is given by the binomial coefficient $d_n=C_{4}^n$. The $n=0$ and $n=4$ states have the highest energy, $E=0$. All eight $n=1$ and $n=3$ states have equal energy, $E=-3U$ for the Heisenberg case, $E=-2U$ in the XY case, and $E=-U$ in the Ising case. In all three cases, the $n=2$ sector contains the lowest energy states. In total, there are six states in the $n=2$ sector, given by:
\begin{equation}
    \Phi = \begin{pmatrix}
    \ket{0;\uparrow\downarrow} & \ket{\uparrow;\uparrow} & \ket{\uparrow;\downarrow} & \ket{\downarrow;\uparrow} & \ket{\downarrow;\downarrow} & \ket{\uparrow\downarrow;0}
    \end{pmatrix}^T\,,
\end{equation}
The matrix elements of $H_4$ in this sector are

\begin{equation}
    \mathcal{H}_{4,\text{Heisenberg}} = -U
    \begin{pmatrix}
    6 & 0 & 0 & 0 & 0 & -6\\
    0 & 4 & 0 & 0 & 0 & 0\\
    0 & 0 & 2 & 2 & 0 & 0\\ 
    0 & 0 & 2 & 2 & 0 & 0\\
    0 & 0 & 0 & 0 & 4 & 0\\
    -6 & 0 & 0 & 0 & 0 & 6
    \end{pmatrix},
\end{equation}
\begin{equation}
    \mathcal{H}_{4,\text{XY}} = -U
    \begin{pmatrix}
    4 & 0 & 0 & 0 & 0 & -4\\
    0 & 4 & 0 & 0 & 0 & 0\\
    0 & 0 & 0 & 0 & 0 & 0\\ 
    0 & 0 & 0 & 0 & 0 & 0\\
    0 & 0 & 0 & 0 & 4 & 0\\
    -4 & 0 & 0 & 0 & 0 & 4
    \end{pmatrix},
\end{equation}
\begin{equation}
    \mathcal{H}_{4,\text{Ising}} = -U
    \begin{pmatrix}
    2 & 0 & 0 & 0 & 0 & -2\\
    0 & 0 & 0 & 0 & 0 & 0\\
    0 & 0 & 2 & 2 & 0 & 0\\ 
    0 & 0 & 2 & 2 & 0 & 0\\
    0 & 0 & 0 & 0 & 0 & 0\\
    -2 & 0 & 0 & 0 & 0 & 2
    \end{pmatrix}.
\end{equation}
The single-site energy spectrum of each case is shown in Fig. \ref{fig:single_site}. The states for the filling $n=2$ are shown in red, whereas the states for the other filling sectors are shown in blue. Both the Heisenberg and XY Hamiltonians have a unique ground state given by $\ket{a_{1}} \equiv \frac{1}{\sqrt{2}}\left(\ket{\uparrow\downarrow;0}-\ket{0;\uparrow\downarrow}\right)$, with energies $-12U$ and $-8U$ respectively. For the Ising Hamiltonian, on the other hand, the lowest-energy state $-4U$ is two-fold degenerate, consisting of both $\ket{a_1}$ and $\ket{a_2}\equiv \frac{1}{\sqrt{2}}\left(\ket{\uparrow;\downarrow}+\ket{\downarrow;\uparrow}\right)$. A general ground state can therefore be written as:

\begin{equation} \label{spinor}
    \ket{\varphi}=\alpha \ket{a_1}+\beta\ket{a_2}\,,
\end{equation}
where $|\alpha|^2+|\beta|^2=1$. This degeneracy is a result of the commutation relation $[H_{4,\text{Ising}},M^z_i]=0$, which does not hold in the Heisenberg or XY cases. Because $M^z_i\ket{a_1}=-2\ket{a_2}$ and $M^z_i\ket{a_2}=-2\ket{a_1}$, the combinations $\frac{1}{\sqrt{2}}(\ket{a_1}- \ket{a_2})$ and $\frac{1}{\sqrt{2}}(\ket{a_1}+\ket{a_2})$ are eigenstates of $M_i^z$ with eigenvalues $\pm 2$. These combinations can thus be interpreted as the two possible polarizations of the inter-band Ising magnetization. Therefore, in the two-dimensional space spanned by the spinor in Eq. (\ref{spinor}), the inter-band magnetization is represented by the $\tau^x$ pseudospin.

This analysis shows that, in the strong-coupling limit, the Heisenberg and XY cases display only a featureless quantum paramagnetic state. On the other hand, the Ising case has a residual SU(2) degeneracy related to the inter-band magnetization degree of freedom. Perturbative interactions between the local inter-band magnetizations, which we will study in the next subsection, may result in non-trivial ground states. Therefore, within our model, the Ising case is the closest analogue of the single-band Hubbard model, whose strong-coupling limit is characterized by perturbative interactions between local (intra-band) spins.

\subsection{Effective Hamiltonian of the Ising case}

We proceed to discuss the effects of the kinetic terms given by $H_2$. For the Heisenberg and XY cases, the many-body ground state is unique and gapped, given by $\ket{\Psi_0} = \Pi_{i=1}^{L^2} \ket{a_1}$. Note that this is a quantum paramagnetic state with $\langle \vec{M}_i e^{i\bQ\cdot \br_i}\rangle=0$ for arbitrary wavevector $\bQ$.
In the Ising case, however, the lowest energy state on each site is a spinor, analogous to the physical electron spin in the one-band Hubbard model. As a result, the ground state has an $\text{SU(2)}^{L^2}$-degeneracy. Perturbations due to the kinetic terms can lift this degeneracy, leading to non-trivial correlations in the Hilbert space spanned by the local spinors. 

The energetics of low-energy excitations can be studied using an effective Hamiltonian approach~\cite{SachdevQPT} in terms of the pseudospins $\tau^\mu$ previously introduced. Recall that, in the basis of Eq. (\ref{spinor}), inter-band magnetic order corresponds to a finite expectation value for $\tau^x$. For the Ising Hamiltonian we find, up to second order in peturbation theory,
\begin{equation}
\begin{split}
    H_{\text{eff},ss'} & \approx -4UL^2\delta_{ss'} + \bra{\Psi_{0,s}} H_2 \ket{\Psi_{0,s'} } \\
    & + \sum_{n\neq 0,t} \frac{ \bra{\Psi_{0,s}} H_2 \ket{\Psi_{n,t}}\bra{\Psi_{n,t}}H_2\ket{\Psi_{0,s'}} }{E_0-E_{n}},
\end{split}
\end{equation}
where $\ket{\Psi_{0,s}} = \Pi_{i=1}^{L^2}\ket{\varphi_{i,s}}$, and $\ket{\varphi_{i,s}}= \alpha_{i,s}\ket{a_1}+\beta_{i,s}\ket{a_2}$  is a configuration of the local spinor. $\ket{\Psi_{n,t}}$ denotes an excited state, having e.g. 1 electron on site $i_1$ and 3 electrons on site $i_2$.

The contribution from the single-particle onsite $\mu^{c,d}$ terms to the effective Hamiltonian is given by
\begin{equation}
	\begin{split}
    	H_{\text{eff}}^{(\mu)} &\approx -\sum_{i} \left[(\mu^c+\mu^d)+\frac{1}{8U}(\mu^c-\mu^d)^2\right]\tau_{i}^0 \\
	 & - \sum_{i} \frac{1}{8U}(\mu^c-\mu^d)^2 \tau^z_{i},
    \end{split}
\end{equation}

Note that only the difference in the onsite energies between the two bands, but not the chemical potential, gives non-trivial energetics. It corresponds to a transverse field, since magnetic order is given by $\tau^x$.  This difference between onsite energies is a tuning parameter absent from the strong-coupling limit of one-band Hamiltonians, such as the Hubbard model. 

The hopping parameter $t_{ij}$ generates spinor correlations between sites $\{i,j\}$. To second order, this leads to the superexchange interactions:
\begin{equation}
     H_{\text{eff}}^{(t)} \approx  \sum_{:ij:}\left(- \frac{(t_{ij}^{c})^2+(t_{ij}^{d})^2}{6U}\tau_{i}^0\tau^0_{j} + \frac{t_{ij}^ct_{ij}^d}{3U}\tau_{i}^x\tau_{j}^x\right),
\end{equation}
where $:ij:$ denotes an ordered pair of sites, i.e., permuting the two indices does not lead to a new term in the Hamiltonian. We note that the superexchange interaction is of an Ising-type, as expected from the fact that the model lacks spin-rotational symmetry. Furthermore, the signs of the exchange interactions depend on the relative signs between the hopping parameters of the two bands. Hence, equal (opposite) signs favor antiferromagnetic (ferromagnetic) alignment of the inter-band magnetization. 
To summarize, in the strong coupling limit, the two-band electronic model with Ising interaction maps onto the generalized transverse-field Ising model containing longer-range exchange interactions:
\begin{equation} \label{eq:TFIM}
    H_{\text{eff}} \approx \sum_{:ij:} J_{ij} \tau_{i}^x\tau^x_{j} - h\sum_{i} \tau_{i}^z \,,
\end{equation}
where $h_i\equiv (\mu_{i}^c-\mu_{i}^d)^2/8U$ and $J_{ij}\equiv t_{ij}^ct_{ij}^d/3U$. Note that we have omitted an overall shift of the ground state energy. 

\section{The transverse-field $J_1$-$J_2$ Ising model}\label{sec:tfim}

\begin{figure}
    \centering
    \includegraphics[width=0.8\linewidth]{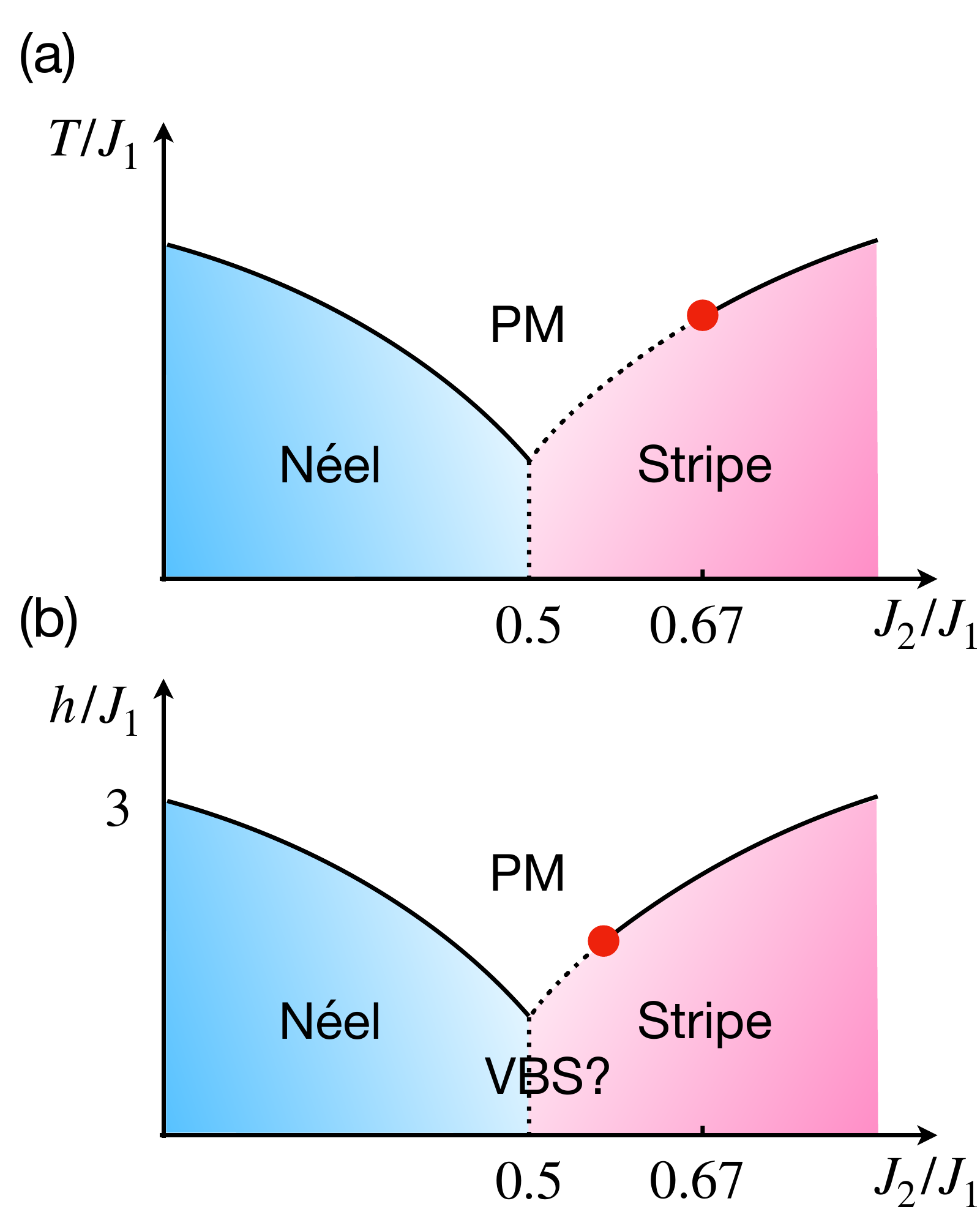}
    \caption{(a) Schematic phase diagram at zero transverse field ($h=0$), and (b) Schematic phase diagram at zero-temperature ($T=0$) of the transverse-field $J_1$-$J_2$ Ising model for antiferromagnetic exchange interactions.
    Solid (dashed) lines correspond to continuous (first order) phase boundaries. PM denotes the paramagnetic phase and VBS denotes a possible string valence-bond solid.}
    \label{fig:TFIM}
\end{figure}

Let us focus on the case where $H_2$ has hopping parameters defined up to next-nearest-neighbor bonds, yielding the band dispersions $\varepsilon^{i}(\bk) = -\mu^i + 2t^i_1(\cos k_x + \cos k_y) + 4t^i_{2}\cos k_x\cos k_y$. According to Eq.~\eqref{eq:TFIM}, this leads to the square-lattice transverse-field $J_1$-$J_2$ Ising model, with
\begin{equation}
    \begin{split}
        J_1 &= \frac{t^c_1t^d_1}{3U}\,,\\ 
        J_2 &= \frac{t^c_2t^d_2}{3U}\,,\\
        h &= \frac{(\mu^c-\mu^d)^2}{8U}\,.
    \end{split}
\end{equation}
This model has been studied extensively in the recent literature~\cite{Jin2013,Kato2015,Sadrzadeh2016,Bobak2018,Kellerman2019,Oitmaa2020}, and here we give a brief overview of the proposed phase diagrams of the frustrated case, where $J_2 > 0$. Because the model is invariant upon changing the signs of $J_1$ or $h$, hereafter we focus on the case of $J_1, h >0$. For $h=0$, the classical phase diagram is rather well-established, and is schematically shown in Fig. \ref{fig:TFIM}(a) based on the Monte Carlo results of Ref.~\onlinecite{Jin2013}. For $J_2<J_1/2$, the ground state is a N\'eel antiferromagnet described by the order parameter $\Delta_{\text{N}} = L^{-2}\sum_{i}\langle \tau^x_i e^{i\bQ_\text{N}\cdot\br_i}\rangle$, where $\bQ_\text{N}\equiv(\pi,\pi)$ is the ordering wave-vector (note that the ground state would be a ferromagnet if $J_1 < 0$). On the other hand, for $J_2>J_1/2$, the ground state is a striped antiferromagnet (regardless of the sign of $J_1$), with an order parameter $\Delta_{\text{S}_n} =L^{-2}\sum_{i} \langle \tau^x_i e^{i\bQ_{\text{S}_n}\cdot\br_i}\rangle$, where $\bQ_{\text{S}_1} = (\pi,0)$ and $\bQ_{\text{S}_2} = (0,\pi)$. The transition between the N\'eel and the stripe phases is first-order, as expected from the fact that they break different symmetries. 

Upon increasing temperature, the N\'eel to paramagnetic (PM) transition is second-order and belongs to the 2D-Ising universality class -- except possibly in a narrow region near $J_2 = J_1/2$, where the transition may become first order \cite{Jin2013}. As for the stripe ground state, it is important to note that it is four-fold degenerate, since there are two stripe ordering wave-vectors and two Ising spin polarizations. In the range $1/2 < J_2/J_1 \lessapprox 0.67$, the stripe-PM transition is first-order. However, for $J_2/J_1 \gtrapprox 0.67$, the stripe-PM transition is second-order and described by the 2D four-state clock model \cite{Jin2013}. A special property of this model is that it only has weak universality, in the sense that only the anomalous critical exponent $\eta$ (and consequently the $\delta$ exponent) is universal, while the other critical exponents are non-universal \cite{Nelson1977}. In the particular case of the classical $J_1$-$J_2$ Ising model, it was shown in Ref. \onlinecite{Jin2013} that the non-universal critical exponents as a function of the ratio $J_2/J_1$ map onto the critical exponents of another model that belongs to the four-state clock weak universality: the Ashkin-Teller model. In particular, for $J_2/J_1 \approx 0.67$, the critical exponents are those of the four-state Potts model, whereas for $J_2/J_1 \rightarrow \infty$, they are those of the 2D Ising model.

The quantum phase diagram ($T=0$) of the model remains widely debated, although some properties seem to be consistent across different methods \cite{Kato2015,Sadrzadeh2016,Bobak2018,Kellerman2019,Oitmaa2020}. In Fig.~\ref{fig:TFIM}(b), we show a schematic candidate phase diagram based on these works. As in the classical case, the N\'eel state is realized for $J_2/J_1 < 1/2$ and the stripe state is realized for $J_2/J_1 > 1/2$. For zero transverse field, the transition at $J_2 = J_1/2$ is first-order, like in the classical phase diagram. As $h$ increases, however, the situation is less clear. For instance, Ref. \onlinecite{Sadrzadeh2016} reported the onset of a string valence-bond solid for finite transverse fields. The quantum phase transition from the N\'eel to the PM state is believed to be second-order and in the 3D Ising universality class \cite{Oitmaa2020} -- although some methods report a regime of first-order transition near $J_2/J_1 = 1/2$ \cite{Bobak2018}. The stripe-PM quantum transition seems to display a tricritical point separating a first-order transition line from a second-order transition line -- similarly to the classical case \cite{Kellerman2019}. Interestingly, the position of the quantum tricritical point seems to be closer to the degeneracy point than the classical tricritical point \cite{Kellerman2019,Oitmaa2020}, i.e. $(J_2 / J_1)_{\mathrm{q}} < (J_2 / J_1)_{\mathrm{cl}} \approx 0.67$, as illustrated schematically in Fig.~\ref{fig:TFIM}(a) and (b) by the red dots.  The nature of the second-order quantum stripe-PM transition remains unclear. Because the classical transition is described by the four-state clock model, it is natural to expect that the quantum transition should be described by the quantum version of the same model. Based on the recent results of Ref. \onlinecite{Patil2020} on the quantum $q$-state clock model, this transition is expected to belong to the 3D XY universality class. However, Ref. \onlinecite{Oitmaa2020} found that the quantum stripe-PM transition has non-universal critical exponents.

This rich landscape of possible ground states of the two-band model with dominant inter-band interactions in the strong-coupling regime contrasts with the simple N\'eel state obtained for the single-band Hubbard model. We now show that the $(h/J_1,\, J_2/J_1)$ phase diagram of Fig. \ref{fig:TFIM}(b) can in principle be traversed with reasonable band structure parameters, contrasting the strong-coupling and weak-coupling behaviors of the model.

\begin{figure}
    \centering
    \includegraphics[width=\linewidth]{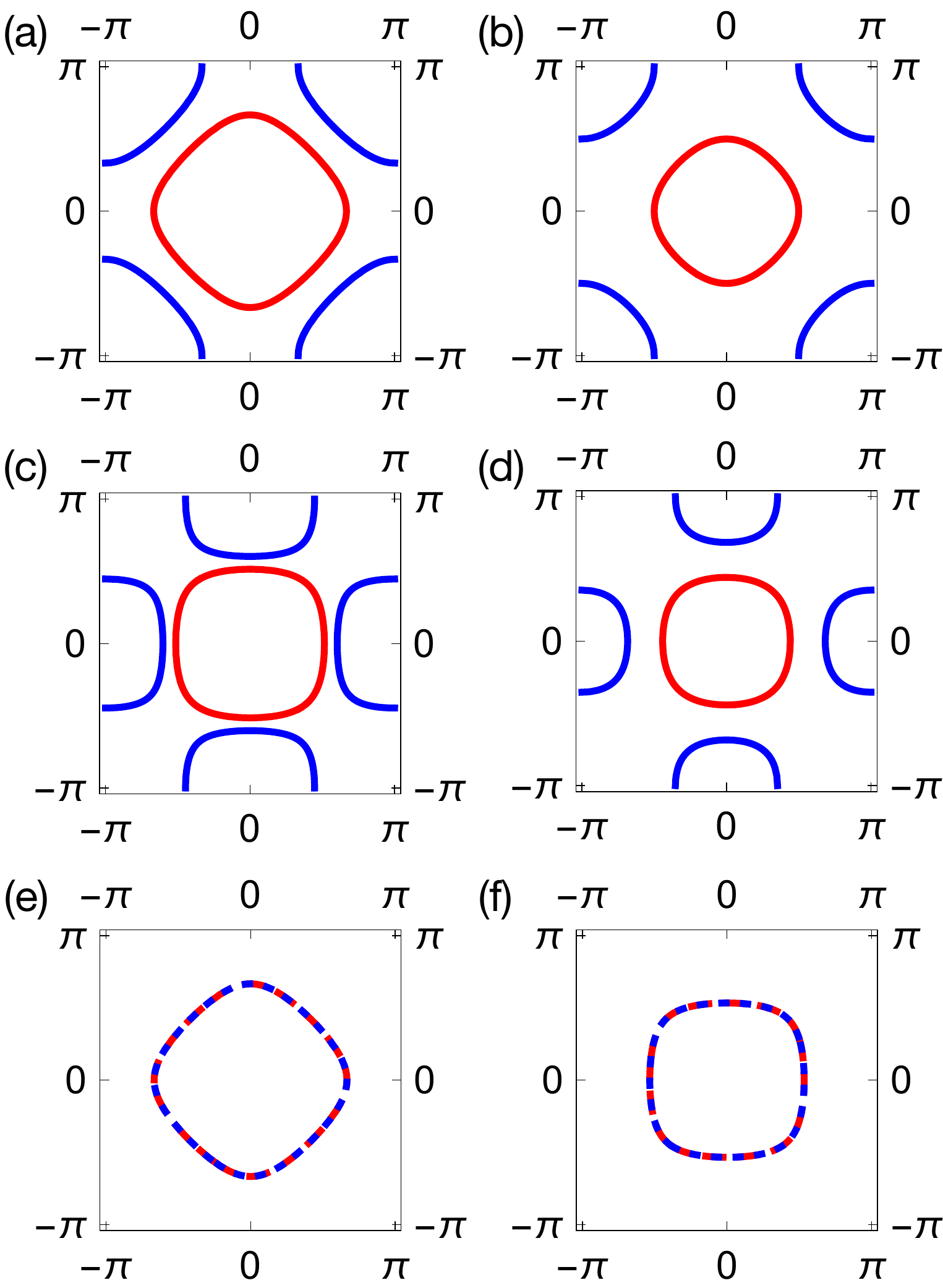}
    \caption{(a-d) Two types of non-interacting Fermi surfaces featuring one hole-like pocket centered at $\Gamma$ and one electron-like pocket centered at either (a,b) $\bQ_\text{N}=(\pi,\pi)$ or (c,d) $\bQ_\text{S}=(\pi,0)/(0,\pi)$. (e-f) Two identical copies of Fermi surfaces with a larger hopping parameter for (e) the nearest-neighbor hopping and (f) the next-nearest-neighbor hopping, respectively. The band parameters are given in Table~\ref{tab:parameters}.}
    \label{fig:2bandFS}
\end{figure}

\begin{table}[]
    \centering
    \begin{tabular}{c|cccccc|cc|c}
        & $\mu^c$ & $t^c_1$ & $t^c_2$ & $\mu^d$ & $t^d_1$ & $t^d_2$ & $J_2/J_1$ & $h/J_1$ & {G.S. ($U \gg t$)} \\
        \hline
        \ref{fig:2bandFS}(a) & $t$ & $t$ & $0$ & $-t$ & $t$ & $0$ & $0$ & $3/2$ &  N\'eel\\
        \ref{fig:2bandFS}(b) & $2t$ & $t$ & $0$ & $-2t$ & $t$ & $0$ & $0$ & $6$ & PM \\
        \ref{fig:2bandFS}(c) & $t$ & $0.6t$ & $0.8t$ & $-t$ & $0.2t$ & $t$ & $20/3$ &  $25/6$ & Stripe \\
        \ref{fig:2bandFS}(d) & $2t$ & $0.6t$ & $0.8t$ & $-2t$ & $0.2t$ & $t$ & $20/3$ &  $50/3$ & PM \\
        \ref{fig:2bandFS}(e) & $t$ & $t$ & $0$ & $t$ & $t$ & $0$ & $0$ & $0$ & N\'eel \\
        \ref{fig:2bandFS}(f) & $t$ & $0.8t$ & $t$ & $t$ & $0.8t$ & $t$ & $25/16$ &  $0$ & Stripe 
    \end{tabular}
    \caption{Parameters of the dispersions shown in Fig.~\ref{fig:2bandFS} and discussed in the text, and their respective ground states (G.S.) at strong coupling.}
    \label{tab:parameters}
\end{table}

A commonly studied situation is when the two bands give rise to a hole-like and an electron-like Fermi pocket, as illustrated in Fig.~\ref{fig:2bandFS} (a)-(d). In (a), the $c$-band creates a hole-like Fermi pocket centered at $\Gamma$ in the Brillouin zone, whereas the $d$-band gives rise to an electron-like Fermi pocket of identical size centered at $\bQ_\text{N}$. This is achieved by a dominant nearest-neighbor hopping, as well as opposite onsite energies, $\mu^c=-\mu^d$, of magnitudes comparable to the bandwidth. As shown in Table~\ref{tab:parameters}, the tight-binding parametrization of Fig.~\ref{fig:2bandFS}(a) favors a N{\'e}el ground state in the strong-coupling regime. Upon increasing $(\mu^c - \mu^d)$, which corresponds to shrinking the sizes of the hole-like and electron-like pockets, the value of the effective random field $h/J_1$ increases and moves the system towards the quantum paramagnetic ground state. The corresponding Fermi surface is shown in Fig.~\ref{fig:2bandFS} (b). Therefore, without changing the electronic occupation, it is possible to induce a quantum phase transition from the N\'eel phase to the paramagnetic phase. Physically, changing the band offset may be achieved via pressure. 

It is also interesting to compare the strong-coupling ground state with the weak-coupling one. The parametrization in Fig.~\ref{fig:2bandFS} (a) satisfies the perfect nesting condition ($\varepsilon^{c}_{\bk}=-\varepsilon^d_{\bk+\bQ_\text{N}}$), which leads to a spin-density wave order, even for infinitesimal small $U$, described by the same N\'eel order parameter. Varying $(\mu^c - \mu^d)$ does not spoil the perfect nesting condition, and therefore is not expected to drive a transition to a paramagnetic phase, in contrast to the strong-coupling limit. 

By including sizable next-nearest-neighbor hopping, i.e. increasing $t_2/t_1$, the $d$-band electron-like Fermi pockets become centered at $\bQ_{\text{S}_1} = (\pi,0)$ and $\bQ_{\text{S}_2} = (0,\pi)$, as shown in Fig.~\ref{fig:2bandFS}(c). Here, a stripe state is preferred at strong-coupling, as shown in Table~\ref{tab:parameters}. Similarly to the previous case, by increasing the onsite energy difference -- see Fig.~\ref{fig:2bandFS}(d) -- the system moves towards a quantum paramagnetic state due to the increase in the transverse field value. In the weak-coupling regime, the stripe magnetic transition temperature is suppressed upon making the nesting conditions poorer. Moreover, as discussed in Ref. \onlinecite{Wang2015}, in the weak-coupling regime, strong deviations from perfect nesting can change the magnetic ground state from the stripe phase to a charge-spin density-wave -- i.e. a collinear double-$\bQ$ state consisting of a linear combination of the order parameters $\Delta_{\mathrm{S_1}}$ and $\Delta_{\mathrm{S_2}}$. Therefore, the magnetic ground states in the weak- and strong-coupling regimes may be different.

There is an important difference between the N\'eel state and the stripe state. The fourfold degeneracy of the latter corresponds to two distinct Ising symmetries: one related to the polarization of the Ising-magnetic order and one related to the tetragonal symmetry of the lattice -- since there are two possible stripe directions that lower the tetragonal symmetry to orthorhombic in different ways. The latter is thus associated with nematic order \cite{Fernandes2012,Fradkin2010}, described in terms of the composite order parameter $\Delta_{\mathrm{S_1}}^2 - \Delta_{\mathrm{S_2}}^2$ \cite{Chandra1990}. Such a nematic phase is called a vestigial order of the underlying stripe phase \cite{Fernandes2019,Nie2014}. In the weak-coupling regime, the quantum nematic phase transition is generally expected to be first-order and simultaneous to the stripe one \cite{Fernandes2013}. However, in the strong-coupling regime of the model, both order parameters can onset simultaneously at a \emph{single} quantum critical point, which likely belongs to the 3D XY universality class. This result has important implications for the possibility of nematic and magnetic quantum criticality being realized in iron-based superconductors \cite{Fang2008,Xu2008,Abrahams2011,Fernandes2012}, whose band structure contains hole pockets and electron pockets separated by the wave-vectors $\bQ_{\text{S}_n}$.

The above examples demonstrate that the magnetic ground states at weak and at strong coupling can break the same symmetry, therefore allowing for a smooth connection at moderate coupling strengths. However, this is by no means necessary. To illustrate this point, we consider a simple example of identical tight-binding parameters, $\mu^c=\mu^d$ and $t^c_{1(2)}=t^d_{1(2)}$, as illustrated in Figs.~\ref{fig:2bandFS}(e) and (f) . In this case, the transverse field vanishes, and the ground state in the strong-coupling regime is either a N\'eel state or a stripe state, depending on the ratio $t_2/t_1$ (see Table~\ref{tab:parameters}). This is in stark contrast to what one would expect from a weak-coupling approach, since this case features two identical Fermi surfaces, which are far from satisfying any nesting condition. 

\section{Conclusions}\label{sec:conclusions}
In this work, we generalized the strong-coupling expansion of the single-band Hubbard model, as pioneered by Phil Anderson \cite{Anderson_superexchange}, to a two-band electronic model with dominant inter-band repulsion. While the former maps onto the Heisenberg model, the latter maps onto the transverse-field Ising model with extended exchange interactions. In particular, the onsite energy difference between the two electronic bands gives rise to a transverse field, while the Ising superexchange interactions arise from virtual hopping processes involving nearest-neighbors and next-nearest-neighbors. Importantly, by fixing the ratio between the inter-band repulsive interactions, this electronic model can be simulated using sign-problem free QMC for arbitrary electronic filling and hopping parameters. In contrast, the Hubbard model, where intra-band repulsion dominates, is only sign-problem-free at half-filling and for bipartite kinetic Hamiltonians.

We also showed how the rich $(h/J_1, \, J_2/J_1)$ phase diagram of the transverse-field $J_1$-$J_2$ model can be probed by appropriately tuning the microscopic band parameters over a reasonable range of values. This opens an interesting avenue for future sign-problem-free QMC studies to explore the impact of different types of strong-coupling ground states on the emergence of superconductivity and other emergent phenomena at intermediate interaction strengths.

\begin{acknowledgements}
We thank Andrey Chubukov, Anders Sandvik, Yoni Schattner, Jun Takahashi, and Oskar Vafek for fruitful discussions. XW acknowledges financial support from National MagLab, which is funded by the National Science Foundation (DMR-1644779) and the state of Florida. MHC acknowledges financial support from the Villum foundation. EB was supported by the European Research Council (ERC) under grant HQMAT (grant no. 817799), the US-Israel Binational Science Foundation (BSF), and the Minerva foundation. RMF is supported by the U.S. Department of Energy, Office of Science, Basic Energy Sciences, Materials Science and Engineering Division, under Award No. DE-SC0020045. RMF also acknowledges partial support from the Research Corporation for Science Advancement via the Cottrell Scholar Award.
\end{acknowledgements}

\bibliography{references}

\end{document}